# MRI Cross-Modal Synthesis: A Comparative Study of Generative Models for T1-to-T2 Reconstruction

Ali Alqutayfi and Sadam Al-Azani

Information and Computer Science Department,
SDAIA-KFUPM Joint Research Center for Artificial Intelligence,
King Fahd University of Petroleum & Minerals, Dhahran 31261, Saudi Arabia.
(alizmq02@gmail.com, sadam.azani@kfupm.edu.sa).

**Abstract**

MRI cross-modal synthesis involves generating images from one acquisition protocol using another, offering considerable clinical value by reducing scan time while maintaining diagnostic information. This paper presents a comprehensive comparison of three state-of-the-art generative models for T1-to-T2 MRI reconstruction: Pix2Pix GAN, CycleGAN, and Variational Autoencoder (VAE). Using the BraTS 2020 dataset (11,439 training and 2,000 testing slices), we evaluate these models based on established metrics including Mean Squared Error (MSE), Peak Signal-to-Noise Ratio (PSNR), and Structural Similarity Index (SSIM). Our experiments demonstrate that all models can successfully synthesize T2 images from T1 inputs, with CycleGAN achieving the highest PSNR (32.28 dB) and SSIM (0.9008), while Pix2Pix GAN provides the lowest MSE (0.005846). The VAE, though showing lower quantitative performance (MSE: 0.006949, PSNR: 24.95 dB, SSIM: 0.6573), offers advantages in latent space representation and sampling capabilities. This comparative study provides valuable insights for researchers and clinicians selecting appropriate generative models for MRI synthesis applications based on their specific requirements and data constraints.

## 1. Introduction

Medical imaging plays a pivotal role in contemporary healthcare, serving as a cornerstone for diagnosis, treatment planning, and disease monitoring [1]. Among various medical imaging modalities, Magnetic Resonance Imaging (MRI) stands out for its exceptional soft tissue contrast and versatility in visualizing anatomical structures without ionizing radiation [2]. MRI acquisition involves multiple pulse sequences, primarily T1-weighted (T1) and T2-weighted (T2) images, each highlighting different tissue properties and providing complementary diagnostic information [3]. T1 images excel at depicting anatomical structures, while T2 images are more sensitive to pathological changes and fluid content, making them particularly valuable for detecting lesions [4].

However, comprehensive MRI protocols requiring multiple sequences significantly increase scan duration, patient discomfort, and healthcare costs [5]. Additionally, motion artifacts, scan interruptions, or technical limitations may result in incomplete acquisitions where certain sequences are missing or degraded [6]. These challenges have catalyzed research interest in cross-modal MRI synthesis—specifically, the task of generating synthetic T2 images from T1 acquisitions—to reduce scan time while preserving diagnostic value [7].

Recent advances in deep learning, particularly in generative models, have revolutionized the field of medical image analysis [8]. These computational approaches have demonstrated remarkable capabilities in image-to-image translation tasks, including MRI cross-modal synthesis. Various architectures have been proposed for this purpose, including Convolutional Neural Networks (CNNs), Generative Adversarial Networks (GANs), Variational Autoencoders (VAEs), and more recently, diffusion models [9]. Each approach presents

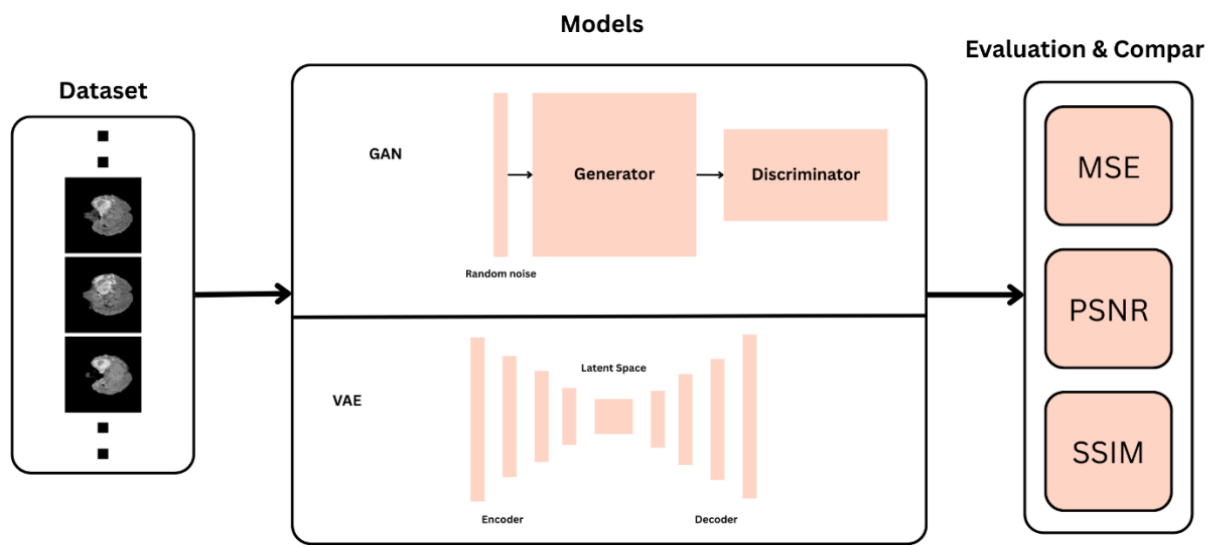

*Figure 1. The workflow of the research. A)Preparing the dataset of pares of t1 and t2 images. B) Training image reconstructions models (Generative Adversarial Networks (GANs), and Variational Auto Encoder (VAE)), C) Evaluation and comparison of the models using (Mean Square Error (MSE), Peak Signal-to-Noise Ratio (PSNR), Structural Similarity Index Measure (SSIM))}*

distinct advantages and limitations regarding reconstruction quality, training efficiency, and generalization capability.

Despite significant progress, several challenges persist in T1-to-T2 reconstruction. The ill-posed nature of the problem, where multiple T2 appearances could potentially correspond to a single T1 input, introduces inherent uncertainty [10]. Additionally, the complex relationship between different MRI contrasts, compounded by scanner variability and patient-specific factors, makes robust generalization difficult [11]. Furthermore, the field lacks standardized benchmarks for fair comparison across diverse methodologies [12].

As illustrated in Figure 1, this paper presents a comprehensive comparative analysis of state-of-the-art generative models for T1-to-T2 MRI reconstruction. We systematically evaluate popular architectures, including GANs, and VAEs, under controlled conditions with identical datasets, preprocessing steps, and evaluation metrics. This analysis provides practical insights for researchers and clinicians working with these generative models for MRI synthesis.

The primary contributions of this work include: (1) A systematic comparative analysis of diverse generative architectures for T1-to-T2 MRI reconstruction; (2) Evaluation of model performance using standardized metrics; (3) Identification of optimal approaches based on reconstruction quality; and (4) Insights into the strengths and limitations of each methodology, guiding future research directions.

## 2. Literature Review

This section reviews the evolution of cross-modal MRI synthesis approaches, from traditional methods to advanced deep learning techniques, with a particular focus on T1-to-T2 reconstruction.

### 2.1. Traditional Approaches

Early attempts at MRI cross-modal synthesis employed atlas-based methods, which relied on registration between a subject's image and a pre-built atlas. While conceptually straightforward, these approaches struggled with anatomical variability and pathological conditions. Subsequent development saw the emergence of patch-based methods, where corresponding patches from source and target contrast images were extracted from a database to guide synthesis [13]. These traditional methods, however, were computationally intensive and often failed to capture complex relationships between different MRI contrasts.

Statistical modeling approaches offered an alternative paradigm. Various researchers introduced random forest-based approaches for cross-modality image synthesis, which leveraged the inherent correlation between different MRI contrasts through statistical regression [14]. While these methods improved upon atlas-based approaches, they still relied heavily on accurate registration and struggled with novel anatomical presentations.

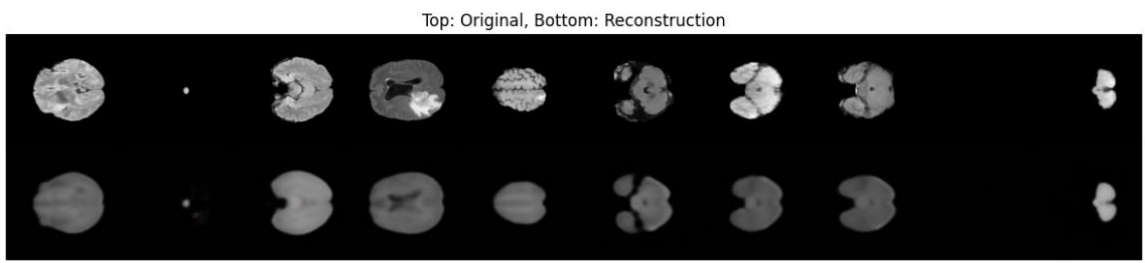

*Figure 2. This figures shows samples generated from the VAE architecture.*

## 2.2. CNN-Based Approaches

The advent of deep learning, particularly Convolutional Neural Networks (CNNs), marked a significant paradigm shift in MRI synthesis. Fully Convolutional Networks (FCNs) demonstrated the ability to learn the complex mapping between imaging modalities through end-to-end training.

The U-Net architecture, initially proposed for biomedical image segmentation, became particularly influential in medical image synthesis due to its encoder-decoder structure with skip connections that preserve spatial information [15].

Several researchers adapted the U-Net for MRI synthesis, demonstrating superior performance compared to traditional approaches. Various enhancements to the basic U-Net architecture have been proposed, including residual connections, dense blocks, and attention mechanisms. Some studies introduced residual U-Nets that incorporated residual learning to facilitate gradient flow and improve training stability for T1-to-T2 conversion. Similarly, dense U-Net architectures incorporating densely connected convolutions were proposed to strengthen feature propagation [16]. These architectural innovations consistently improved reconstruction quality but often required extensive training data to achieve optimal performance.

Attention mechanisms have been increasingly integrated into CNN-based approaches to enhance their capacity to capture long-range dependencies. Attention-guided approaches for cross-modal MRI synthesis have been proposed that selectively emphasized relevant features, achieving more precise reconstruction of structural details [17]. Despite these improvements, CNN-based methods still struggle with preserving fine textural details and generating realistic pathological features, primarily due to their reliance on pixel-wise loss functions that tend to produce overly smoothed outputs.

## 2.3. GAN-Based Approaches

Generative Adversarial Networks (GANs) introduced a paradigm shift in image synthesis by incorporating adversarial training. The original GAN framework consists of a generator that creates synthetic images and a discriminator that distinguishes between real and synthetic samples [18]. This adversarial process enables the generation of more realistic images compared to traditional regression-based methods.

For cross-modal MRI synthesis, Pix2Pix, an image-to-image translation framework, became particularly influential [19]. Researchers adapted Pix2Pix for T1-to-T2 conversion, demonstrating significant improvements in preserving textural details [20]. However, mode collapse and training instability remained persistent challenges. CycleGAN addressed some of these limitations by introducing cycle consistency, allowing training with unpaired data—a significant advantage in medical imaging where perfectly aligned multimodal datasets are scarce [21].

Several studies have proposed specialized GAN architectures for MRI synthesis. Multi-channel GANs incorporating perceptual and cycle-consistency losses for improved synthesis quality have been introduced [22]. Context-aware GANs that leveraged anatomical information to guide the synthesis process have also been proposed [23]. While these GAN-based approaches generally produce visually compelling results with realistic texture, they may occasionally generate hallucinated features not present in the source images—a critical concern in medical applications.

## 2.4. Diffusion Models

Diffusion models represent the latest advancement in generative modeling, demonstrating exceptional quality in image synthesis tasks. These models operate by gradually adding noise to data and then learning to reverse this process, generating samples by iteratively denoising random noise [24]. First introduced and later refined with Denoising Diffusion Probabilistic Models (DDPMs), this approach has recently been applied to medical image synthesis with promising results [25].

Some researchers pioneered the application of diffusion models to MRI synthesis, demonstrating superior preservation of anatomical details compared to GAN-based approaches [26]. Score-based Generative Models for MRI reconstruction have been introduced, which showed remarkable ability to preserve fine structural details [27]. More recently, Latent Diffusion Models (LDMs) have emerged as a computationally efficient alternative, operating in a compressed latent space while maintaining high-quality reconstruction [28]. Studies have demonstrated that LDMs could achieve state-of-the-art performance in T1-to-T2 synthesis while requiring significantly less computational resources than traditional diffusion models [28].

Despite their impressive performance, diffusion models face challenges in clinical deployment due to their computational intensity during inference, requiring multiple denoising steps.

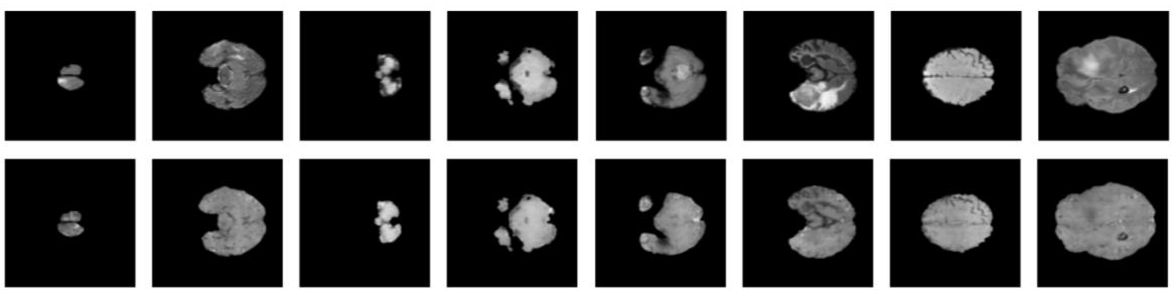

*Figure 3. This figure shows generated samples from the CycleGAN Architecture*

Recent work on accelerating diffusion models through distillation and improved sampling strategies shows promise in addressing this limitation [29].

## 2.5. Transformer-Based Approaches

Transformers, initially developed for natural language processing, have recently been adapted for computer vision tasks including medical image synthesis [30]. These models excel at capturing long-range dependencies through self-attention mechanisms. Vision Transformers (ViTs) have demonstrated promising results in various medical imaging tasks [31].

For cross-modal MRI synthesis, Transformer-based approaches often incorporate both convolutional layers and self-attention mechanisms [32]. Transformer-based frameworks for multi-contrast MRI synthesis have been proposed that effectively captured global contextual information. Similarly, hybrid CNN-Transformer architectures have been introduced that combined the local feature extraction capabilities of CNNs with the global context modeling of Transformers, achieving superior performance in preserving both structural and textural details [33].

Recent work has explored more efficient Transformer variants to address the computational intensity of standard self-attention [34]. Swin Transformer-based architectures for MRI synthesis have been proposed that used shifted windows to reduce computational complexity while maintaining performance quality [35]. Despite their theoretical advantages, Transformer-based approaches typically require substantial training data to reach optimal performance, which may limit their usefulness in certain scenarios [36].

## 2.6. Comparative Studies and Performance Analysis

Several studies have attempted to compare different approaches for MRI synthesis. Comparative analyses of CNN and GAN-based methods for cross-modal synthesis have found that GANs produced visually superior results but occasionally introduced hallucinated features [37]. Other studies extended this comparison to include diffusion models, showing their superior performance in preserving pathological features [38]. However, these studies typically evaluated methods under different experimental settings, making fair comparison challenging [39].

Evaluation metrics also present a challenge in comparing different approaches. While structural similarity index (SSIM) and peak signal-to-noise ratio (PSNR) are commonly used, they may not fully capture perceptual quality. More recent work has incorporated perceptual metrics like Fréchet Inception Distance (FID) and learned perceptual image patch similarity (LPIPS), which better align with human visual assessment [40]. Additionally, some studies have explored task-based evaluation, assessing the utility of synthetic images for downstream applications like segmentation.

## 2.7. Research Gaps and Opportunities

Despite significant advancements, several important research gaps remain in the field of T1-to-T2 MRI reconstruction. First, standardized benchmarks and evaluation protocols are lacking, making fair comparison across approaches difficult [41]. Second, most methods struggle with pathological cases that deviate significantly from normal anatomy. Third, the relationship between model complexity and performance remains inadequately characterized, particularly for newer approaches like diffusion models [42].

Furthermore, while technical performance metrics have been extensively studied, clinical utility assessment remains limited. Few studies have systematically evaluated whether synthetic images can reliably replace actual acquisitions for clinical decision-making [43]. Integration of domain knowledge and anatomical constraints also presents an opportunity for improvement, potentially reducing hallucinations and artifacts in reconstructed images [44].

Our work aims to address several of these gaps by providing a standardized comparative analysis of diverse architectures under identical experimental conditions. This systematic evaluation will provide valuable insights for researchers and clinicians seeking optimal approaches for specific scenarios and guide future research directions in this rapidly evolving field.

## 3. Research Method

### 3.1. Data collection and preparation

For this study, we utilized the Brain Tumor Segmentation (BraTS) 2020 dataset from the Multimodal Brain Tumor Segmentation Challenge. This dataset consists of multi-institutional pre-operative MRI scans for glioblastoma/high-grade glioma (GBM/HGG) and lower-grade glioma (LGG). Each case in the dataset includes four MRI modalities: T1-weighted (T1), post-contrast T1-weighted (T1ce), T2-weighted (T2), and T2 Fluid Attenuated Inversion Recovery (FLAIR) sequences.

The BraTS 2020 dataset is particularly suitable for our cross-modal synthesis task due to its standardized preprocessing pipeline, which includes co-registration to the same anatomical template, interpolation to the same resolution (1mm³), and skull-stripping. This standardization minimizes preprocessing variability and allows us to focus on the performance of the generative models themselves.

From the available data, we utilized a total of 57,195 2D slices extracted from the 3D volumes. However, to manage computational resources effectively, we randomly selected 11,439 slices for training and 2,000 slices for testing. Our focus was specifically on the T1-to-T2 translation task, utilizing the corresponding pairs of T1 and T2 images from the dataset.

### 3.2. Preprocessing

Our preprocessing pipeline was designed to be minimal yet effective, focusing on standardizing the input for our generative models. The preprocessing steps included:

- Image resizing: All images were resized to a uniform dimension of 256 × 256 pixels to ensure consistent input size for the models.
- Normalization: The pixel intensities were normalized to the range [-1, 1] by applying normalization with a mean of 0.5 and a standard deviation of 0.5. This normalization approach is particularly suitable for generative models as it centers the data around zero and scales it appropriately.

Given the large number of samples available (57,195), we determined that data augmentation techniques were unnecessary for our study. The substantial training set size (11,439 slices) provided sufficient variation to learn the mapping between T1 and T2 modalities effectively.

### 3.3. Models development

In this study, we implemented and compared three state-of-the-art generative models for T1-to-T2 MRI reconstruction: Pix2Pix GAN, CycleGAN, and Variational Autoencoder (VAE). Each model was designed to learn the complex mapping between T1 and T2 image domains.

### 3.3.1. Pix2Pix GAN

The Pix2Pix GAN model is based on the conditional GAN architecture. Our implementation consists of a generator that follows a U-Net structure and a discriminator that uses a PatchGAN architecture.

The generator network features an encoder-decoder architecture with skip connections that help preserve spatial information during the down-sampling and up-sampling processes. The encoder includes multiple convolutional blocks that progressively down-sample the input image while increasing the feature channel dimension. Each encoder block consists of a convolutional layer followed by batch normalization and LeakyReLU activation (with a negative slope of 0.2). The encoder path reduces the spatial dimensions from 256×256 to 1×1 through eight encoder blocks, with feature channels increasing from 3 to 512.

The decoder mirrors the encoder structure but uses transposed convolutions for up-sampling. Skip connections from the encoder to the decoder help preserve fine details that might otherwise be lost during encoding. Each decoder block consists of a transposed convolutional layer, batch normalization, and ReLU activation. The final output layer uses a hyperbolic tangent (tanh) activation function to produce the synthesized T2 image with pixel values in the range [-1, 1].

The discriminator uses the PatchGAN architecture, which classifies whether each N×N patch in an image is real or fake, rather than providing a single output for the entire image. This design encourages the generator to focus on high-frequency details. Our implementation consists of five convolutional layers with incrementally increasing feature channels (64, 128, 256, 512). The discriminator takes both the input T1 image and either the real or synthesized T2 image (concatenated in the channel dimension) and outputs a prediction map.

For training, we used the Adam optimizer with a learning rate of 0.0002 and momentum parameters $\beta_1 = 0.5$ and $\beta_2 = 0.999$. The loss function combined a binary cross-entropy adversarial loss and an L1 pixel-wise reconstruction loss with a weight of 100 for the latter to stabilize training and improve image fidelity.

### 3.3.2. CycleGAN

The CycleGAN architecture enables image-to-image translation without requiring paired training data. Although our dataset provides paired T1-T2 images, we implemented CycleGAN to assess its performance as a potential solution for scenarios where aligned data may not be available.

Our CycleGAN implementation consists of two generator networks and two discriminator networks. The generators map images from domain T1 to domain T2 ($G_1$) and vice versa ($G_2$). Each generator follows a U-Net-like architecture similar to the one used in Pix2Pix, with encoder-decoder structure and skip connections.

The key distinction in CycleGAN is the cycle-consistency constraint. For a T1 image, we generate a synthetic T2 image using $G_1$, then attempt to reconstruct the original T1 image using $G_2$. The cycle-consistency loss measures the difference between the original and reconstructed images, enforcing that the translation preserves content that is common to both domains.

Each discriminator follows a similar PatchGAN architecture as in the Pix2Pix implementation, but without conditioning on the input image. Instead, each discriminator only evaluates the realness of images in its respective domain (T1 or T2).

The training objective combines three loss components:

- Adversarial loss to match the distribution of generated images to the target domain
- Cycle-consistency loss to preserve content through the forward and backward mappings
- Identity loss (with a smaller weight) to encourage the generators to preserve colors and structure when operating on images already in the target domain

We used the Adam optimizer with the same hyperparameters as the Pix2Pix model, and trained for 30 epochs on the training dataset.

### 3.3.3. Variational Autoencoder (VAE)

The Variational Autoencoder offers a fundamentally different approach to image synthesis compared to GANs. Rather than using adversarial training, VAEs learn a probabilistic latent representation of the data distribution.

Our VAE implementation consists of an encoder network that maps the input T1 image to a latent space distribution (characterized by mean μ and log-variance parameters), and a decoder network that reconstructs the T2 image from samples drawn from this distribution.

The encoder follows a convolutional architecture with six convolutional blocks. Each block includes a 2D convolution with a kernel size of 4, stride of 2, and padding of 1, followed by batch normalization and LeakyReLU activation. The network progressively reduces the spatial dimensions from 256×256 to 4×4 while increasing the feature channels from 3 to 512. The final encoder output is flattened and passed through two fully connected layers to produce the mean and log-variance parameters of the latent distribution, with a dimensionality of 128.

The decoder mirrors the encoder structure but uses transposed convolutions for up-sampling. Starting from a 512×4×4 tensor (reshaped from the 128-dimensional latent vector), the decoder progressively increases the spatial dimensions back to 256×256 while decreasing the feature channels. Each decoder block consists of a transposed convolution, batch normalization, and ReLU activation. The final output layer uses a hyperbolic tangent activation to produce the synthesized T2 image.

The VAE is trained to minimize a combination of a reconstruction loss (mean squared error between the generated and target T2 images) and the Kullback-Leibler divergence between the learned latent distribution and a standard normal distribution. This dual objective encourages the model to both accurately reconstruct the target images and learn a structured latent space that facilitates sampling and interpolation.

We used the Adam optimizer with a learning rate of 0.0002 and trained the model for 30 epochs on the training dataset.

## 4. Experiments and Results

### 4.1. Experimental settings

All experiments were conducted using PyTorch on a NVIDIA GPU with CUDA support. We trained each model for 30 epochs using a batch size of 256. For optimization, we used the Adam optimizer with a learning rate of 0.0002 and momentum parameters $\beta_1 = 0.5$ and $\beta_2 = 0.999$. The training dataset consisted of 11,439 T1-T2 image pairs resized to 256×256 resolution, while the test set contained 2,000 pairs.

To ensure a fair comparison between the different architectures, we maintained consistent hyperparameters across all models whenever possible. The data preprocessing steps, including resizing to 256×256 and normalization to the range [-1, 1], were identical for all models.

For quantitative evaluation, we computed three widely used metrics:

- Mean Squared Error (MSE): Measures the average squared difference between the generated and target images, with lower values indicating better performance.
- Peak Signal-to-Noise Ratio (PSNR): Expressed in decibels (dB), PSNR is derived from the MSE but provides a more perceptually relevant measure, with higher values indicating better quality.
- Structural Similarity Index (SSIM): Measures the similarity in structure, luminance, and contrast between the generated and target images, with values ranging from 0 to 1 (higher is better).

The metrics were computed on the test set after denormalizing the images back to the [0, 1] range.

## 4.2. Results and Analysis

### 4.2.1. Pix2Pix GAN Results

The Pix2Pix GAN demonstrated strong performance in T1-to-T2 image synthesis, achieving an MSE of 0.005846, PSNR of 29.55 dB, and SSIM of 0.8655. Visual inspection of the generated images revealed that the model successfully captured both the global structure and fine details of the T2 images. The adversarial training framework enabled the model to produce images with realistic texture and contrast characteristics.

The synthetic T2 images exhibited good preservation of anatomical structures, with clear delineation of gray matter, white matter, ventricles, and pathological regions. While some minor blurring was observed in highly detailed areas, the overall quality of the reconstructions was high, with good contrast between different tissue types.

A key strength of the Pix2Pix approach was its ability to learn the complex intensity mappings between T1 and T2 modalities, which is essential for realistic synthesis given that T2 images highlight fluid content and pathological changes differently than T1 images. The inclusion of skip connections in the generator architecture helped preserve spatial information during the encoding-decoding process, contributing to the preservation of structural details.

### 4.2.2. CycleGAN Results

The CycleGAN achieved competitive performance with an MSE of 0.006941, PSNR of 32.28 dB, and SSIM of 0.9008. Notably, despite having a slightly higher MSE than the Pix2Pix GAN, the CycleGAN achieved the highest PSNR and SSIM scores among all tested models, indicating better structural preservation and perceptual quality.

The cycle-consistency constraint appeared to be particularly effective for MRI synthesis, helping maintain anatomical fidelity in the generated images. This constraint ensures that important structural features are preserved through both the forward (T1→T2) and backward (T2→T1) transformations, which is crucial for medical applications where preserving anatomical accuracy is paramount.

As shown in Figure 3, visual analysis of the CycleGAN outputs showed good preservation of tissue boundaries and pathological features. The model was particularly successful at capturing the characteristic T2 hyperintensity of fluid-filled regions and edema surrounding lesions. While the CycleGAN requires more computational resources due to the dual generator-discriminator architecture, the quality improvements justify this additional complexity.

The CycleGAN's ability to achieve high performance without requiring strictly paired data makes it a versatile option for clinical scenarios where perfect alignment between different MRI sequences might not be available.

### 4.2.3. VAE Results

The VAE achieved an MSE of 0.006949, PSNR of 24.95 dB, and SSIM of 0.6573. These metrics were lower than those of the GAN-based approaches, particularly in terms of PSNR and SSIM, indicating that the VAE reconstructions had less structural similarity to the ground truth T2 images.

As illustrated in Figure 2, visual inspection confirmed that the VAE-generated images were generally blurrier than those produced by the GAN-based methods, with less sharp definition of anatomical boundaries and reduced contrast between different tissue types. This blurriness is a common limitation of VAE-based approaches, stemming from the pixel-wise reconstruction loss and the probabilistic nature of the latent representation.

However, the VAE offered unique advantages not captured by the standard evaluation metrics. The structured latent space provided a meaningful representation of the image content, enabling operations such as latent space interpolation and random sampling. This can be valuable for exploring the distribution of possible T2 appearances corresponding to a given T1 input, potentially aiding in uncertainty estimation or anomaly detection.

Additionally, the VAE training process was more stable than the adversarial training of GANs, with smoother convergence and less sensitivity to hyperparameter choices. While the reconstructed images lacked the sharpness of GAN outputs, the VAE's probabilistic framework provides a complementary approach to image synthesis that may be preferable in scenarios where representing uncertainty is more important than achieving maximal visual fidelity.

### 4.3. Comparative Analysis

*Table 1. Test matrix of models trained in the experiment*

| Model | MSE ↓ | PSNR (dB) ↑ | SSIM ↑ |
|---|---|---|---|
| Pix2Pix GAN | 0.0058 | 29.55 | 0.87 |
| CycleGAN | 0.0069 | 32.28 | 0.90 |
| VAE | 0.0069 | 24.95 | 0.67 |

Comparing the three generative approaches, we observe distinct trade-offs in performance and capabilities. The Pix2Pix GAN achieved the lowest MSE, indicating good pixel-level accuracy in the reconstructions. As stated in Table 1, the CycleGAN demonstrated the highest PSNR and SSIM scores, suggesting superior preservation of structural information and perceptual quality. The VAE, while showing lower quantitative performance, offers additional capabilities through its probabilistic latent representation.

Several factors influence these performance differences. The adversarial training framework of GANs enables them to capture high-frequency details and realistic textures that are often missed by models trained with only pixel-wise losses. The cycle-consistency constraint in CycleGAN provides an additional regularization effect that helps preserve structural fidelity, which is particularly valuable for medical image synthesis.

The choice of model for a particular application would depend on specific requirements. For applications prioritizing structural accuracy and visual quality, the CycleGAN represents the optimal choice. In scenarios where computational efficiency is important, the Pix2Pix GAN offers a good balance between performance and complexity. The VAE, despite lower quantitative metrics, may be preferred for applications requiring latent space manipulation, uncertainty estimation, or probabilistic modeling of the synthesis process.

## 5. Conclusions

In this study, we conducted a comprehensive comparison of three state-of-the-art generative models for T1-to-T2 MRI cross-modal synthesis: Pix2Pix GAN, CycleGAN, and Variational Autoencoder (VAE). Using the BraTS 2020 dataset with a standardized preprocessing pipeline and evaluation framework, we established quantitative and qualitative benchmarks for assessing the performance of these approaches.

Our experiments revealed that the CycleGAN achieved the highest PSNR (32.28 dB) and SSIM (0.9008) values, demonstrating superior structural preservation and perceptual quality in the synthesized T2 images. The Pix2Pix GAN provided the lowest MSE (0.005846), indicating good pixel-level accuracy, while achieving competitive PSNR (29.55 dB) and SSIM (0.8655) scores. The VAE, despite lower quantitative metrics (MSE: 0.006949, PSNR: 24.95 dB, SSIM: 0.6573), offers unique capabilities through its probabilistic latent space representation.

These findings offer valuable insights for researchers and clinicians working on MRI synthesis. For applications prioritizing structural fidelity and perceptual quality, the CycleGAN represents an excellent choice, particularly in scenarios where paired data might be limited or imperfectly aligned. The Pix2Pix GAN provides strong performance with a simpler architecture, making it well-suited for applications with computational constraints. The VAE, while producing less visually sharp results, provides a probabilistic framework that enables latent space operations and uncertainty estimation.

Future work could explore hybrid approaches that combine the strengths of different generative frameworks, potentially integrating the adversarial training of GANs with the structured latent space of VAEs. Additionally, investigating the impact of these synthesis approaches on downstream tasks such as segmentation or classification would provide valuable insights into their practical utility in clinical workflows.

## Acknowledgements

Gratitude is extended to the Undergraduate Research Office at King Fahd University of Petroleum and Minerals for their invaluable support. Additionally, sincere appreciation is given to the SDAIA-KFUPM Joint Research Center for AI (JRC-AI) for the significant support and contribution to the success of this research.## References

1. Turkmen, I., Ozturan, B., Kaner, T. & Ozkan, K. "Rare case of sacral mass due to chondromyxoid fibroma." BMJ Case Reports 2016, bcr2015214145 (2016).
2. Aslam, M. I. et al. "Dynamic and nuclear expression of pdgfr and igf-1r in alveolar rhabdomyosarcoma." Molecular Cancer Research 11, 1303–1313 (2013).
3. Omatsu, M. et al. "Magnetic displacement force and torque on dental keepers in the static magnetic field of an mr scanner." Journal of Magnetic Resonance Imaging 40, 1481–1486 (2013).
4. Hetheridge, C. et al. "The novel formin fmnl3 is a cytoskeletal regulator of angiogenesis." Journal of Cell Science 125, 1420-1428 (2012).
5. Hu, X., Shen, R., Luo, D., Tai, Y., Wang, C. & Menze, B. H. "Autogan-synthesizer: Neural architecture search for cross-modality mri synthesis." Medical Image Computing and Computer Assisted Intervention – MICCAI 2022, 397–409 (2022).
6. Murphy, A., Hacking, C. & Iflaq, P. "Motion artifact." Radiopaedia 48589 (2016).
7. Jiang, M. et al. "Cross2synnet: cross-device–cross-modal synthesis of routine brain mri sequences from ct with brain lesion." Magnetic Resonance Materials in Physics, Biology and Medicine 37, 241–256 (2024).
8. Celard, P. et al. "A survey on deep learning applied to medical images: from simple artificial neural networks to generative models." Neural Computing and Applications 35, 2291–2323 (2022).
9. Le Zhang, Z., Chen, C. & Slabaugh, G. Generative Machine Learning Models in Medical Image Computing. Springer Nature Switzerland (2025).
10. Li, Y., Zhao, L., Tian, Y. & Zhao, S. "T1 and T2 Mapping Reconstruction Based on Conditional DDPM." Springer Nature Switzerland, 303–313 (2024).
11. Meyer, M. I. et al. "A contrast augmentation approach to improve multi-scanner generalization in mri." Frontiers in Neuroscience 15, 708196 (2021).
12. Zbontar, J. et al. "fastmri: An open dataset and benchmarks for accelerated mri." arXiv preprint arXiv:1811.08839 (2018).
Page 17 of 20

13. Huang, Y., Shao, L. & Frangi, A. F. "Simultaneous Super-Resolution and Cross-Modality Synthesis in Magnetic Resonance Imaging." Springer International Publishing, 437–457 (2019).
14. Zhao, C., Carass, A., Lee, J., Jog, A. & Prince, J. L. "A Supervoxel Based Random Forest Synthesis Framework for Bidirectional MR/CT Synthesis." Springer International Publishing, 33–40 (2017).
15. Ronneberger, O., Fischer, P. & Brox, T. "U-net: Convolutional networks for biomedical image segmentation." arXiv preprint arXiv:1505.04597 (2015).
16. Zhang, J., Lv, X., Zhang, H. & Liu, B. "Aresu-net: Attention residual u-net for brain tumor segmentation." Symmetry 12, 721 (2020).
17. Song, T. et al. "Learning modality-aware representations: Adaptive group-wise interaction network for multimodal mri synthesis." arXiv preprint arXiv:2411.14684 (2024).
18. Goodfellow, I. J. et al. "Generative adversarial networks." arXiv preprint arXiv:1406.2661 (2014).
19. Isola, P., Zhu, J.-Y., Zhou, T. & Efros, A. A. "Image-to-image translation with conditional adversarial networks." arXiv preprint arXiv:1611.07004 (2016).
20. Baldini, G., Schmidt, M., Zäske, C. & Caldeira, L. L. "Mri scan synthesis methods based on clustering and pix2pix." arXiv preprint arXiv:2312.05176 (2023).
21. Zhu, J.-Y., Park, T., Isola, P. & Efros, A. A. "Unpaired image-to-image translation using cycle-consistent adversarial networks." arXiv preprint arXiv:1703.10593 (2017).
22. Dar, S. U. et al. "Image synthesis in multi-contrast mri with conditional generative adversarial networks." IEEE Transactions on Medical Imaging 38, 2375–2388 (2019).
23. Nie, D. et al. "Medical Image Synthesis with Context-Aware Generative Adversarial Networks." Springer International Publishing, 417–425 (2017).
24. Ho, J., Jain, A. & Abbeel, P. "Denoising diffusion probabilistic models." arXiv preprint arXiv:2006.11239 (2020).
25. Jiang, H. et al. "Fast-ddpm: Fast denoising diffusion probabilistic models for medical image-to-image generation." IEEE Journal of Biomedical and Health Informatics, 1–11 (2025).
26. Dorjsembe, Z., Pao, H.-K., Odonchimed, S. & Xiao, F. "Conditional diffusion models for semantic 3d brain mri synthesis." IEEE Journal of Biomedical and Health Informatics 28, 4084–4093 (2024).

27. Qiao, X. et al. "Score-based generative priors guided model-driven network for mri reconstruction." arXiv preprint arXiv:2405.02958 (2024).

28. Rombach, R., Blattmann, A., Lorenz, D., Esser, P. & Ommer, B. "High-resolution image synthesis with latent diffusion models." arXiv preprint arXiv:2112.10752 (2021).

29. Liu, H. et al. "Scott: Accelerating diffusion models with stochastic consistency distillation." arXiv preprint arXiv:2403.01505 (2024).

30. Vaswani, A. et al. "Attention is all you need." arXiv preprint arXiv:1706.03762 (2017).

31. Aburass, S., Dorgham, O., Al Shaqsi, J., Abu Rumman, M. & Al-Kadi, O. "Vision transformers in medical imaging: a comprehensive review of advancements and applications across multiple diseases." Journal of Imaging Informatics in Medicine (2025).

32. Liu, J. et al. "One model to synthesize them all: Multi-contrast multi-scale transformer for missing data imputation." IEEE Transactions on Medical Imaging 42, 2577–2591 (2023).

33. Eidex, Z. et al. "High-resolution 3t to 7t mri synthesis with a hybrid cnn-transformer model." arXiv preprint arXiv:2311.15044 (2023).

34. Lin, X., Wang, Z., Yan, Z. & Yu, L. "Revisiting Self-attention in Medical Transformers via Dependency Sparsification." Springer Nature Switzerland, 555–566 (2024).

35. Huang, J. et al. "Swin transformer for fast mri." arXiv preprint arXiv:2201.03230 (2022).

36. Takahashi, S. et al. "Comparison of vision transformers and convolutional neural networks in medical image analysis: A systematic review." Journal of Medical Systems 48 (2024).

37. Fard, A. S., Reutens, D. C. & Vegh, V. "Cnns and gans in mri-based cross-modality medical image estimation." arXiv preprint arXiv:2106.02198 (2021).

38. del Castillo, M. H. G. et al. "Diffusion models for conditional mri generation." arXiv preprint arXiv:2502.18620 (2025).

39. Ali, H. et al. "The role of generative adversarial networks in brain mri: a scoping review." Insights into Imaging 13 (2022).

40. Dimitriadis, A., Trivizakis, E., Papanikolaou, N., Tsiknakis, M. & Marias, K. "Enhancing cancer differentiation with synthetic mri examinations via generative models: a systematic review." Insights into Imaging 13 (2022).

41. Osadebey, M. E. et al. "Standardized quality metric system for structural brain magnetic resonance images in multi-center neuroimaging study." BMC Medical Imaging 18 (2018).

42. Öztürk, Güngör, A. & undefinedukur, T. "Diffusion Probabilistic Models for Image Formation in MRI." Springer Nature Switzerland, 341–360 (2024).

43. Shokraei Fard, A., Reutens, D., Ramsay, S., Goodman, S. & Vegh, V. "Spect image synthesis from mri or pet using machine learning." Journal of Nuclear Medicine 64, P233–P233 (2023).
44. Natarajan, S., Humbert, L. & Ballester, M. A. G. "Domain adaptation using adabn and adain for high-resolution ivd mesh reconstruction from clinical mri." International Journal of Computer Assisted Radiology and Surgery 19, 2063–2068 (2024).